\documentclass[conference]{IEEEtran}
\IEEEoverridecommandlockouts

\usepackage[]{graphicx}    
\newcommand{\ignore}[1]{}  
\usepackage{url}
\usepackage{amsmath}
\usepackage{float}
\usepackage{subcaption}
\usepackage{graphicx}
\usepackage{tikz}
\usepackage{algorithm}
\usepackage{algorithmic}
\usepackage{color}
\usetikzlibrary{shapes.arrows}
\usetikzlibrary{decorations.markings}
\tikzstyle{midarrow}=[decoration={markings, mark=at position 0.5 with {\arrow{stealth}}}, postaction={decorate}]
\usetikzlibrary{arrows.meta, positioning}

\usepackage{placeins} 
\usepackage{amssymb}
\newcommand{\NN}{\mathbb{N}}
\newcommand{\RR}{\mathbb{R}}

\graphicspath{{figures/}}

\def\BibTeX{{\rm B\kern-.05em{\sc i\kern-.025em b}\kern-.08em
    T\kern-.1667em\lower.7ex\hbox{E}\kern-.125emX}}
\begin{document}
\title{Quality of Service-Constrained Online Routing in High-Throughput~Satellites}

\author{\IEEEauthorblockN{Olivier Bélanger\IEEEauthorrefmark{1}\IEEEauthorrefmark{2}, Olfa Ben Yahia\IEEEauthorrefmark{1}\IEEEauthorrefmark{2}, Stéphane Martel\IEEEauthorrefmark{3}, Antoine Lesage-Landry\IEEEauthorrefmark{1}\IEEEauthorrefmark{2}, Gunes Karabulut Kurt\IEEEauthorrefmark{1}}

\IEEEauthorblockA{\IEEEauthorrefmark{1}Department of Electrical Engineering, Polytechnique Montréal, Montréal, Qc H3T 1J4}
\IEEEauthorblockA{\IEEEauthorrefmark{2}GERAD, Montréal, Qc H3T 1N8} 
\IEEEauthorblockA{\IEEEauthorrefmark{3}MDA, Sainte-Anne-de-Bellevue, Qc H9X 3R2} 

\textit{\{olivier.belanger, olfa.ben-yahia, antoine.lesage-landry, gunes.kurt\}}@polymtl.ca, stephane.martel@mda.space}

\maketitle

\thispagestyle{plain}
\pagestyle{plain}

\begin{abstract}
High-throughput satellites (HTSs) outpace traditional satellites due to their multi-beam transmission. The rise of low Earth orbit mega constellations amplifies HTS data rate demands to terabits/second with acceptable latency. This surge in data rate necessitates multiple modems, often exceeding single device capabilities. Consequently, satellites employ several processors, forming a complex packet-switch network. This can lead to potential internal congestion and challenges in adhering to strict quality of service (QoS) constraints. While significant research exists on constellation-level routing, a literature gap remains on the internal routing within a single HTS. The intricacy of this internal network architecture presents a significant challenge to achieve high data rates. 

This paper introduces an online optimal flow allocation and scheduling method for HTSs. The problem is presented as a multi-commodity flow instance with different priority data streams. An initial full time horizon model is proposed as a benchmark. We apply a model predictive control (MPC) approach to enable adaptive routing based on current information and the forecast within the prediction time horizon while allowing for deviation of the latter. Importantly, MPC is inherently suited to handle uncertainty in incoming flows. Our approach minimizes packet loss by optimally and adaptively managing the priority queue schedulers and flow exchanges between satellite processing modules. Central to our method is a routing model focusing on optimal priority scheduling to enhance data rates and maintain QoS. The model's stages are critically evaluated, and results are compared to traditional methods via numerical simulations. Through simulations, our method demonstrates performance nearly on par with the hindsight optimum, showcasing its efficiency and adaptability in addressing satellite communication challenges.
\end{abstract} 

\begin{IEEEkeywords}
High-throughput satellite, model predictive control, routing, scheduling, quality of service.
\end{IEEEkeywords}


\section{Introduction}

Satellites play an increasingly vital role in global communications~\cite{Richharia2010}, connecting vast and remote territories and bridging communication gaps in regions where deploying terrestrial infrastructure for both cellular and fiber optic networks remains challenging~\cite{IrekvistLinder2022}. While remote areas have traditionally relied on satellites for communications and television broadcasting~\cite{pratt2019satellite}, in recent years, these services have expanded to include Internet connectivity~\cite{Graydon2020}. This expansion underlines the immense potential and importance of satellite networks, especially when considering the growth of direct cell-to-satellite connectivity~\cite{10068542}. As numerous users with varied needs directly engage with satellite resources, the implementation of resilient and reliable quality of service (QoS) mechanisms becomes essential~\cite{electronics8060683}. Prioritizing traffic, ensuring data delivery, and managing congestion become crucial in such scenarios.

With the growing reach and appeal of these services, there is a notable shift from traditional single-beam satellites to the more agile multi-beam high-throughput satellites (HTSs)~\cite{LUGLIO2022109352}. The movement of Low Earth Orbit (LEO) satellites adds a layer of uncertainty, necessitating rapid adaptation in satellite network management. Unlike their predecessors, HTSs are engineered to handle high data rates, addressing the digital requirements of modern society~\cite{Zhang_2023}. As global demand for swift internet grows, the need for rapid and QoS-reliable HTSs intensifies. The intricacies of HTSs, coupled with their enhanced capabilities, necessitate specialized components for data management. A pivotal part of this internal architecture is the modem banks, as shown in~\cite{Kuiper2019,yahia2023evolution}. They act as the central processing units, ensuring efficient handling, routing, and distribution of the massive influx of data within the satellite system. The interconnections between modem banks bring forward new routing and load balancing challenges. Properly addressing these challenges is of high importance because it can significantly enhance data throughput and reduce latency. In the high-stakes realm of satellite communication, ensuring these efficiencies directly translates to improved user experience and service reliability.

With the global demand for high-quality communication, enhancing the QoS and expanding satellite accessibility to a broader user base is crucial. On the one hand, a large spectrum of research has explored satellite communication, with substantial literature dedicated to enhancing the performance of both traditional satellites and the more advanced HTSs, as well as the networks they create~\cite{itu-r2019key-elements,9861699,7417202,6934560}. 
On the other hand, load balancing strategies, indispensable for evenly distributing incoming data and maintaining peak performance, have gained attention in works like~\cite{electronics11213611}. These strategies are crucial for ensuring that no single modem bank or processing unit is overwhelmed, leading to potential data losses or delays. Efficient load balancing allows for optimal resource utilization, minimizes response time, and prevents system overload. Its relevance to HTS networks, where uneven traffic could lead to significant bottlenecks and degrade QoS, is highlighted in studies such as ~\cite{9843265}.

However, an important gap emerges when addressing internal routing and load balancing specifically tailored to HTSs. HTSs, given their capability to handle high data rates, inherently require specialized internal routing mechanisms to efficiently manage and distribute the incoming data. The variable and uncertain nature of these incoming flows introduces an additional layer of complexity to the routing challenges. The complexities arise particularly when interconnecting modem banks, amplifying the need for strong routing strategies. While many advancements have been made on the constellation-level (inter-satellite) front~\cite{8002583,9500740}, the challenges inherent to internal HTS routing remain comparatively overlooked. This paper seeks to address this gap, introducing innovative techniques to enhance the internal routing capabilities of HTSs and their overall performance.

Improving internal routing in HTSs not only promises more efficient satellite operations but can also catalyze broader socioeconomic advancements by connecting previously under-served and inaccessible communities in remote regions. This, in turn, can pave the way for greater economic opportunities, for example via improved access to information and education. 

\subsection{Contribution}
This research focuses on addressing the challenges associated with flow allocation and scheduling in HTSs within mega constellations. Our primary contribution lies in integrating model predictive control (MPC) associated with incoming data flows. By incorporating MPC, we can develop a predictive model that allows for online adjustments based on incoming data patterns, leading to optimized routing decisions. This integration aims to improve the overall efficiency and reliability of data transmission within the HTS system. We show numerically that the MPC method performs very close to the optimality set by the \textit{batch optimization with hindsight} method, i.e., when all random information is revealed.

\subsection{Paper Organization}
This paper is organized as follows: Section 2 presents the internal model we developed. In Section 3, we outline our proposed approaches based on MPC. Numerical results are presented in Section 4, and Section 5 offers a conclusion and suggests directions for future work. 

\section{Problem Formulation}

In an HTS, there are multiple processing units called modem banks. We consider a satellite equipped with $M \in \NN$ modem banks, with each bank managing $P \in \NN$ distinct priorities. 

Queues play a crucial role in managing the flow of data packets. Each queue, associated with a specific priority within a modem bank, temporarily holds incoming data packets until they can be processed and routed. Complementing the role of the queues are the scheduler weights, $w_{p}^m$. They determine the fraction of packets of priority $p \in \{ 1,2,...,P\}$ that are selected from each queue from module $m \in \{ 1,2,...,M\}$ to be processed and subsequently transmitted. Traditionally, routing techniques relied on fixed weights, which often didn't accommodate varying traffic demands and patterns effectively. By dynamically adjusting these weights, the satellite can adapt to the incoming flows and the current state of the queues. 

\begin{figure*}[t]
    \centering
    \begin{tikzpicture}[scale=0.8], every node/.style={transform shape}]
        \draw[fill=gray!30] (4,4) rectangle (9,5.5);
        \node[anchor=north] at (5,5.5) {Priority 1}; 
        \draw[fill=gray!30] (4,2.5) rectangle (9,4);
        \node[anchor=north] at (5,4) {Priority 2}; 
        \node[anchor=north] at (6.5,6.1) {Module 1}; %
    
        \draw[fill=gray!30] (4,-1) rectangle (9,0.5);
        \node[anchor=north] at (5,-0.4) {Priority 1}; 
        \draw[fill=gray!30] (4,-2.5) rectangle (9,-1);
        \node[anchor=north] at (5,-1.9) {Priority 2};
        \node[anchor=north] at (6.5,-2.5) {Module 2}; 
    
        \draw[dotted, thick] (0,5) rectangle (1.5,-2);
    
        \fill (0.75,4) circle (0.1);
        \fill (0.75,-1) circle (0.1);
        
        \fill (12,1.5) circle (0.1);     
    
        \fill (12,7) circle (0.1);
        \node[right] at (12.2,7) {Sink 1};
        \fill (12,-4) circle (0.1);  
        \node[right] at (12.2,-4) {Sink 2};
    
        \fill (6.5,-1.75) circle (0.1);  
        \fill (6.5,-0.25) circle (0.1);   
        
        \fill (6.5,3.25) circle (0.1);  
        \fill (6.5,4.75) circle (0.1);   
    
        \draw[midarrow] (0.75,4) -- (6.5,4.75);
        \node[above] at (3,4.5) { \( f^{\text{in}, 1}_1(t) \) };
            
        \draw[midarrow] (0.75,4) -- (6.5,-0.25);
        \node[above] at (3,2.75) { \( f^{\text{in}, 2}_1(t) \) };
        
        \draw[midarrow] (0.75,-1) -- (6.5,3.25);
        \node[above] at (3,1) { \( f^{\text{in}, 1}_2(t) \) };
            
        \draw[midarrow] (0.75,-1) -- (6.5,-1.75);
        \node[above] at (3,-1.25) { \( f^{\text{in}, 2}_2(t) \) };
            
        \draw[midarrow] (6.5,4.75) -- (12,1.5);
        \node[above] at (11,2.5) { \( f^{\text{out}, 1}_1(t) \) };

        \draw[midarrow] (6.5,-0.25) -- (12,1.5);
        \node[above] at (9.5,0.8) { \( f^{\text{out}, 2}_1(t) \) };
        
        \draw[midarrow] (6.5,3.25) -- (12,1.5);
        \node[below] at (9.5,2.3) { \( f^{\text{out}, 1}_2(t) \) };
        
        \draw[midarrow] (6.5,-1.75) -- (12,1.5);
        \node[below] at (11,0.5) { \( f^{\text{out}, 2}_2(t) \) };

        \draw[midarrow] (6.5,4.75) -- (12,7);
        \draw[midarrow] (6.5,-0.25) -- (12,7);
        \draw[midarrow] (6.5,3.25) -- (12,-4);
        \draw[midarrow] (6.5,-1.75) -- (12,-4);
    
        \node[single arrow, draw, fill=blue!20, minimum height=2cm, minimum width=1cm, single arrow head extend=.2cm, single arrow head indent=.1cm] (Arrow1) at (-1.5,4) {};
        \node at (Arrow1.center) {\(F_1\)};
        \node[single arrow, draw, fill=blue!20, minimum height=2cm, minimum width=1cm, single arrow head extend=.2cm, single arrow head indent=.1cm] (Arrow1) at (-1.5,-1) {};
        \node at (Arrow1.center) {\(F_2\)};
        
        \node[single arrow, draw, fill=green!20, minimum height=1cm, minimum width=0.5cm, single arrow head extend=.2cm, single arrow head indent=.1cm] at (13, 1.5) {};
    
    \end{tikzpicture}
    \caption{HTS internal routing model for $M=2$ and $P=2$}
    \label{fig:model}
\end{figure*}
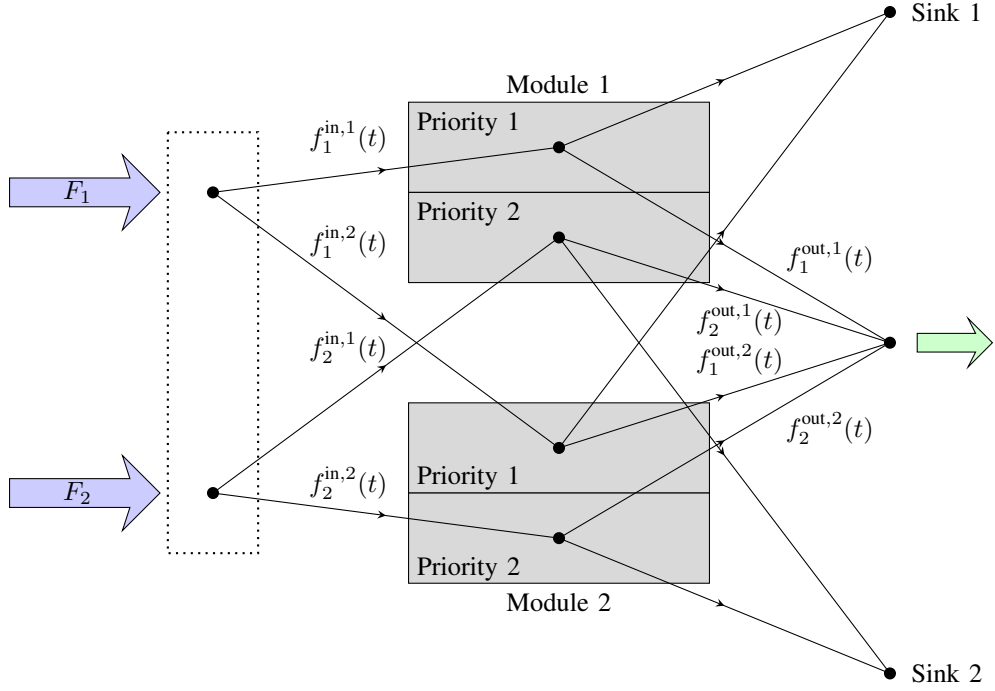

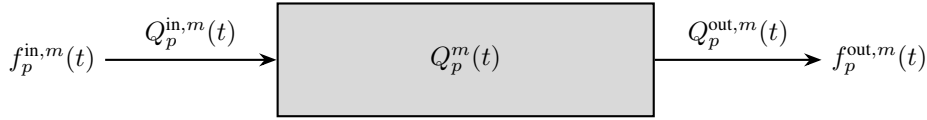
\begin{figure*}[t]
    \centering
    \begin{tikzpicture}[>=Stealth, thick]
      \pgfmathsetmacro{\halfwidth}{5cm / 2}
      
      \node (fin) at (-\halfwidth-3cm,0) {\( f_{p}^{\text{in},m}(t) \)}; 
      \node[rectangle, draw, fill=gray!30, minimum width=5cm, minimum height=1.5cm] (Q) at (0,0) {\( Q_{p}^{m}(t) \)};
      \node (fout) at (\halfwidth+3cm,0) {\( f_{p}^{\text{out},m}(t) \)}; 
    
      \draw[->] (fin) -- (Q) node[midway, above] {\( Q_{p}^{\text{in},m}(t) \)};
      \draw[->] (Q) -- (fout) node[midway, above] {\( Q_{p}^{\text{out},m}(t) \)};
    \end{tikzpicture}
    \caption{HTS internal queuing model}
    \label{fig:queuing_model}
\end{figure*}

We formulate the problem as a multi-commodity flow instance \cite{ahuja1993network}, where the data flows represent the commodities. In this context, the uplink beam serves as the source, and the sinks encompass both the downlink beam and lost packets. Figure \ref{fig:model} presents the simplified case where $M = P = 2$. In Figure \ref{fig:model}, the uplink beam is depicted as two distinct beams for clarity of representation. However, it's essential to understand that these are virtual nodes. There is in reality a single uplink beam that contains all the different priorities. Additionally, Figure \ref{fig:queuing_model} represents an expanded view of the queuing processes within a given module $m$ for a specific priority $p$. We consider the HTS internal problem over a time horizon $T \in \NN$. We discretize the horizon in time steps of duration $\Delta t > 0$ and index them by $t \in \{ 1,2,...,T \}$.

\subsection{HTS Packet Management Model}

The routing scheme of an HTS is characterized by several constraints, as detailed below.

At each time step $t$, the net inflow $f_p^{\text{in}, m}(t) \in \RR^+$ and outflow $f_p^{\text{out},m}(t)\in \RR^+$ of commodity $p \in \{1,2,...,P\}$ at module \mbox{$m \in \{1,2,...,M\}$} are equilibrated by considering the queue's inflow $Q_p^{\text{in},m}(t) \in \RR^+$ and outflow $Q_p^{\text{out},m}(t) \in \RR^+$ balance, $\Delta Q_p^m(t) \in \RR$, and the packet loss ${\mathcal{L}}_p(t) \in \RR^+$:
\begin{equation}
    f_p^{\text{in}, m}(t) - f_p^{\text{out},m}(t) - \Delta Q_p^m(t) - {\mathcal{L}}_p(t) = 0. 
    \label{eq:constraint1}
\end{equation}For every module, the scheduler weight at a time step $t$ represents the percentage of processed packets from queue of priority $p$:
\begin{equation}
    0 \leq w_{p}^m(t) \leq 1.
    \label{eq:constraint2}
\end{equation}For every module, the scheduler weights $w_{p}^m(t)$ across all priorities $p \in \{1,2,...,P\}$ must sum to 1:
\begin{equation}
    \sum_{p =1}^P w_{p}^m(t) = 1.
    \label{eq:constraint3}
\end{equation}
To avoid high disruptions in the system, we introduce a ramping constraint. For a given module and priority, the weight of a scheduler at a time step $t$ is constrained by the maximum deviation from the previous time step:
\begin{equation}
    w_{p}^m(t) - w_{p}^m(t-1) \leq \overline{\Delta w}.
    \label{eq:constraint4}
\end{equation}At each time step $t$, the outflow $f_p^{\text{out},m}(t)$ from module $m$ represents a fraction of packets processed by that module. It is constrained by the scheduler weights normalized over their respective operation periods, $\Delta s > 0$:
\begin{equation}
    f_p^{\text{out},m}(t) \leq \frac{w_p^m(t)}{\Delta s},
    \label{eq:constraint5}
\end{equation}
where $\Delta s$ is a parameter that depends on the HTS.
The cumulative incoming flow for a specific priority $p$, $f_p^{\text{in}}(t)$, for each module $m \in \{1,2,...,M\}$ matches the observed demand $F_p(t) \in \RR^+$:
\begin{equation}
    \sum_{m =1}^M f_p^{\text{in}}(t) = F_p(t). 
    \label{eq:constraint6} 
\end{equation}
Let $Q_p^m(t)$ be the occupancy of a specific queue at time~$t$. The occupancy at the next time step, \mbox{$Q_p^m(t+1)$}, is determined by its current occupancy $Q_p^m(t)$, to which the net incoming (outgoing) packet balance $\Delta Q_p^m(t)$ is added:
\begin{equation}
    Q_p^m(t+1) = Q_p^m(t) + \Delta Q_p^m(t).
    \label{eq:constraint7}
\end{equation}Each queue starts and concludes with a specific, predefined occupancy $Q_0 \in \RR^+$:
\begin{equation}
    Q_p^m(0) = Q_p^m(T-1) = Q_0.
    \label{eq:constraint8} 
\end{equation}Every queue has a total capacity bounded by $\overline{Q} \in \RR^+$, induced by the hardware limitations of the modem banks in  satellite communication subsystems:
\begin{equation}
    \sum_{p =1}^P Q_p^m(t) \leq \overline{Q}.
    \label{eq:constraint9}
\end{equation}All output links have a maximum capacity $\overline{C} \in \RR^+$, representing the transmission bandwidth constraints of the satellite communication channels:
\begin{equation}
    f_p^{\text{out},m}(t)  \leq \overline{C}.
    \label{eq:constraint10}
\end{equation}
Let $F_p(t)$ be the stochastic process modelling the incoming flow of a given priority $p$. We assume that $F_p(t)$ is distributed according to a Poisson process with an average arrival rate of $\lambda_p$ packets per time step ~\cite{medhi2018network}.

Let $k_p$ be the cost incurred when losing a packet of priority $p \in \{ 1,2,...,P\}$. In the batch optimization problem, informed by $F_p (t)$ in hindsight, we seek to minimize the total packet loss cost over all priorities and time steps:
\begin{align}
    &\min_{w_p^m(t), f_p^{\text{in},m}(t)} \sum_{t=1}^T\sum_{p=1}^{P} {\mathcal{L}}_p(t) k_p \label{eq:objective} \\
    &\text{subject to } (\ref{eq:constraint1})-(\ref{eq:constraint10}). \nonumber
\end{align}
In \eqref{eq:objective}, ${\mathcal{L}}_p(t)$ denotes the total number of lost packets for priority $p$ across all modules at time $t$. The incurred cost when a packet of priority $p$ is lost is $k_p$, with higher priorities corresponding to greater costs. The total cost, which we seek to minimize, is the aggregated sum of these costs over all time steps and priorities.

\section{Proposed Approaches}
In this section, we present the proposed approaches, starting with batch optimization, proceeding to MPC, and concluding with alternative methods.
\subsection{Batch Optimization}

In addressing the task of optimizing flow allocation in HTSs, our initial approach is \textit{batch optimization with hindsight} information on the incoming flow. This method processes data from the entire time horizon to derive an optimal solution~\cite{luenberger2008linear}. This means that the incoming flows are predetermined for the whole horizon and processed as a single batch, i.e., there is no underlying uncertainty.

This approach offers several advantages. Access to the complete data set for the full time horizon allows for the generation of the optimal solution by accounting for future information. This broader perspective enables the system to anticipate and adjust for potential bottlenecks or surges in demand. It ensures that the derived solution optimizes flow allocation throughout the entire time horizon.

However, the comprehensive nature of this method, while it is a strength, can also be viewed as a limitation. Requiring complete information means that the system might fail in dynamic environments with sudden demand shifts. In these circumstances, batch optimization lacks the flexibility to respond to these changes. The method can be computationally demanding, especially when faced with longer horizons or larger data sets. This may result in longer processing times, which is unsuitable for real-time applications, especially given the restricted on-board computing capacity of satellites. As a result, we will consider \textit{batch optimization with hindsight} as the gold standard against which we measure our results.

Within the realm of HTSs, the batch optimization framework can certainly be of use for stable, predictable scenarios. However, considering the dynamic environment of satellite communication, mostly characterized by variable demands in our study, an adaptive strategy might be essential. This leads to the exploration of MPC.

\subsection{Model Predictive Control}

MPC is an advanced control technique that uses a model of the system to predict its future behaviour over a given time horizon~\cite{rawlings2017mpc}. These predictions guide the system in computing controls that optimize specific performance criteria, such as reducing packet loss cost in our case, while taking into account constraints on the system's inputs, states, and outputs. 

In the field of HTSs, the dynamic nature of satellite communication and packet arrival poses challenges that traditional optimization approaches may struggle to address. The ever-changing environment, caused by the varying, uncertain demands and movement of the satellites, requires a method that can adapt swiftly. This is where MPC proves valuable. 

The core idea behind applying MPC to HTS is its inherent adaptability. Unlike batch optimization, which needs the full incoming flow information for the entire time horizon prior to computations, MPC adjusts to new information as it becomes available. By doing so, it continually refines its routing and scheduling decisions, ensuring close to optimal response to evolving scenarios. Instead of working with the realized stochastic process, we work with the expected flow only. In practice, this value can be empirically estimated from historical data. The expression of the flow then becomes:
\begin{equation}
    \hat{F_p}(t)  = {\mathbb E}[F_p(t)],
    \label{eq:distro_mean}
\end{equation}
and \eqref{eq:constraint6} is substituted by:
\begin{equation}
    \sum_{m =1}^M f_p^{\text{in}}(t) =  \hat{F_p}(t).
    \label{eq:constraint5_mpc} 
\end{equation}
The formulation of our MPC approach mirrors the constraints and objectives mentioned in the Batch Optimization subsection. However, its execution differs. For every time step, the MPC method evaluates the optimization problem over a moving time window $W < T$. Once the solution is determined, the initial control action corresponding to the current round is implemented in the HTS. Following this decision, the states of the HTS, such as the outflows and queue occupancies, are updated and observed. With this refreshed state information and new flow data, the time window advances while maintaining its fixed duration. The optimization problem is then revisited and solved for the subsequent time step. As each time step progresses, the system continues this iterative process, leveraging both the current state of the system and, when available, revised flow forecasts. The main difference lies in the optimization function and the online implementation. Problem \eqref{eq:objective} becomes, at time $t \in \{1,2,...,T\}$:
\begin{equation}
\begin{aligned}
& \min_{\substack{w_p^m(\tau), f_p^{\text{in},m}(\tau), \\ \tau \in \{ t,t+1,...,t+W \}}} & &\sum_{\tau=t}^{t+W}\sum_{p=1}^{P} {\mathcal{L}}_p(\tau) k_p. \\
& \text{\quad \;\;\;subject to} & & (\ref{eq:constraint1}) - (\ref{eq:constraint5}), (\ref{eq:constraint7}) - (\ref{eq:constraint10}), (\ref{eq:constraint5_mpc})
\label{eq:objective_mpc}
\end{aligned}
\end{equation}
where $w_p^m(t)$ and $f_p^{\text{in}, m}(t)$ are then implemented.

Our experiments, detailed in the subsequent sections, illustrate MPC's ability to tackle uncertain incoming flow in this setting. The comparison with the batch optimization benchmark reveals that while batch optimization might produce theoretically optimal solutions for static scenarios, MPC excels in dynamic, uncertain conditions typical of satellite communications. Not only does MPC achieve near-optimal results, but it also maintains consistent performance under these ever-changing conditions. Additionally, MPC is computationally efficient because it solves a linear problem using, e.g., off-the-shelf solvers, such as Mosek in our case \cite{mosek}. The MPC process for online routing and scheduling in an HTS is summarized in Algorithm \ref{alg:mpc}.

\begin{algorithm}[tb]
\caption{MPC online routing and scheduling process}\label{alg:mpc}
\begin{algorithmic}[1]
    \STATE \textbf{Parameters:} $W$
    \STATE \textbf{Initialization:} Given $\hat{F_p}(t)$ \FOR{$t$ in $\{1,2, \dots,T\}$}
    \STATE \quad Observe the states $Q_p^m(t), \Delta Q_p^m(t), {\mathcal{L}}_p(t)$ and $f_p^{\text{out},m}(t)$
    \STATE \quad Solve \eqref{eq:objective_mpc} considering $\hat{F_p}(t)$
    \STATE \quad Implement $w_p^m(t)$ and $f_p^{\text{in}, m}(t)$
    \ENDFOR
\end{algorithmic}
\end{algorithm}

\subsection{Alternative Benchmarking Methods}

Apart from batch optimization, the first alternative method is termed \textit{static batch optimization with hindsight}. This method utilizes one less degree of freedom by calculating the optimal weights at the initial time step and maintaining them consistently over time. It still uses hindsight information, making it impractical. It is used as a second benchmark for our approach. It can be represented by adding the following constraint to \eqref{eq:objective}:
\begin{equation}
    w_p^m(t) = w_p^m(1) \quad\forall \quad t \in \{2, 3,...,T \}.
    \label{eq:constraint_static_batch}
\end{equation}

Another method used for comparison is the \textit{windowless MPC} approach. In this method, we employ an MPC with a window of size 1. The objective behind this is to ascertain if the added design variable of window length in our original MPC approach is beneficial.

The last comparison method is the \textit{cost-proportional} method, where the following constraint is added to \eqref{eq:objective}:
\begin{equation}
    w_p^m(t) = \frac{k_p}{\sum_{p=1}^P k_p} \quad\forall \quad t, p.
    \label{eq:constraint_prop}
\end{equation}
\section{Numerical Results}
This section presents our experimental setup, describes the numerical comparison methods, and provides the results along with their analysis.
\subsection{Experimental Setup}
The simulations were constructed to emulate a realistic environment for HTSs. The experimental framework is defined over a time horizon of $T = 100$ time steps, and the MPC approach uses a window $W = 10$. The empirical justification for this choice of window size is provided in the Results and Analysis subsection. The satellites are modelled with $M = 16$ modem banks, which can each accommodate $P = 3$ distinct priorities of data packets, e.g., phone calls (high priority, $p=1$), text messages (medium priority, $p=2$), and emails (low priority, $p=3$). The packet loss costs, corresponding to the significance of these priorities, are 10, 4, and 1, respectively.
Importantly, smaller numbers were used for readability. The entire experimental setup is designed to be scalable, allowing for the representation of flow values across various orders of magnitude, such as $10^9$ packets in the context of an HTS.

The system is constrained such that each queue can hold a maximum of $\overline{Q} = 10$ packets, with both the initial and final states of the queues being empty ($Q_0 = 0$) to promote the continuity of the routing strategy. Lastly, the scheduler weights cannot deviate by more than $\overline{\Delta w}$ = 10\% between consecutive time steps.

In this study, the timeline is divided into segments representing progressively higher traffic, characterized by average arrival rates starting at 10 packets per time step and increasing linearly up to 100. To establish an inversely proportional relationship relative to packet priorities, we normalized $\lambda_p$ by the packet loss cost $k_p$.

\subsection{Comparison Methods}

The proposed MPC approach is rigorously tested against other methods to demonstrate its potential benefits and capabilities. The first benchmark is the \textit{batch optimization with hindsight} method. This is the ``gold standard'', as it utilizes ex post information for the total time horizon, optimizing weights and routing incoming flow based on the actual, realized flow for every time step. However, while it offers a theoretical optimum, its applicability is limited as it uses information not readily available when routing decisions are made.

The second benchmark is a variant of the gold standard, known as the \textit{static batch optimization with hindsight} method. It closely resembles its predecessor, with one added constraint: the weights remain constant over time. This emphasizes consistency in allocation decisions and leads to streamlined computations, but at the expense of adaptability, as it removes one degree of freedom. 

The third benchmark, \textit{windowless MPC}, simplifies the standard MPC approach by using a window of only one time step. This eliminates the potential benefit of considering future states, a key advantage of traditional MPC. However, this simplification translates to increased computational efficiency.


Lastly, the \textit{cost proportional} allocation method, in contrast, presents a simpler and more direct approach. Instead of unpacking the complexities of the incoming flows, it focuses on proportional weight allocations based on costs, ensuring rapid decision-making. While it prioritizes computational efficiency over precision, this method remains an important consideration.

The next subsection details the numerical assessment of the MPC method against these four benchmarks, shedding light on the advantages the MPC approach brings to packet routing in HTSs.

\subsection{Results and Analysis}
\label{sec:subsection-title}
Having outlined the experimental framework and benchmarking methods, this subsection presents the quantitative results of our simulations. To ensure consistency in our findings, all simulations were conducted 100 times using a Monte-Carlo approach. These results offer insights into the efficacy of the MPC technique.

\begin{figure}
    \centering
    \begin{subfigure}[b]{0.99\linewidth}
        \includegraphics[width=\linewidth]{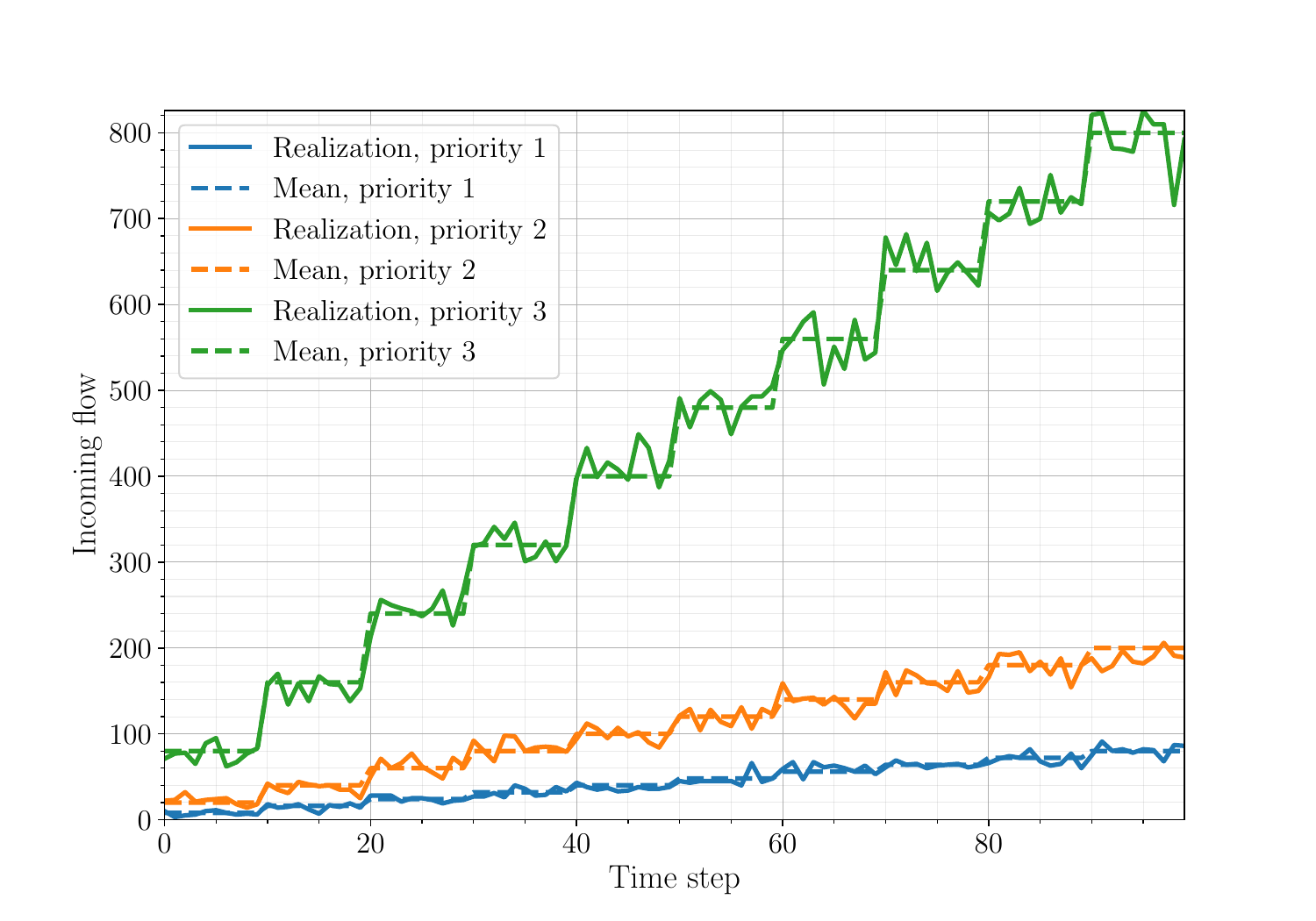}
        \caption{}
        \label{fig:f_in_vs_t_single_run}
    \end{subfigure}
    \begin{subfigure}[b]{0.99\linewidth}
        \includegraphics[width=\linewidth]{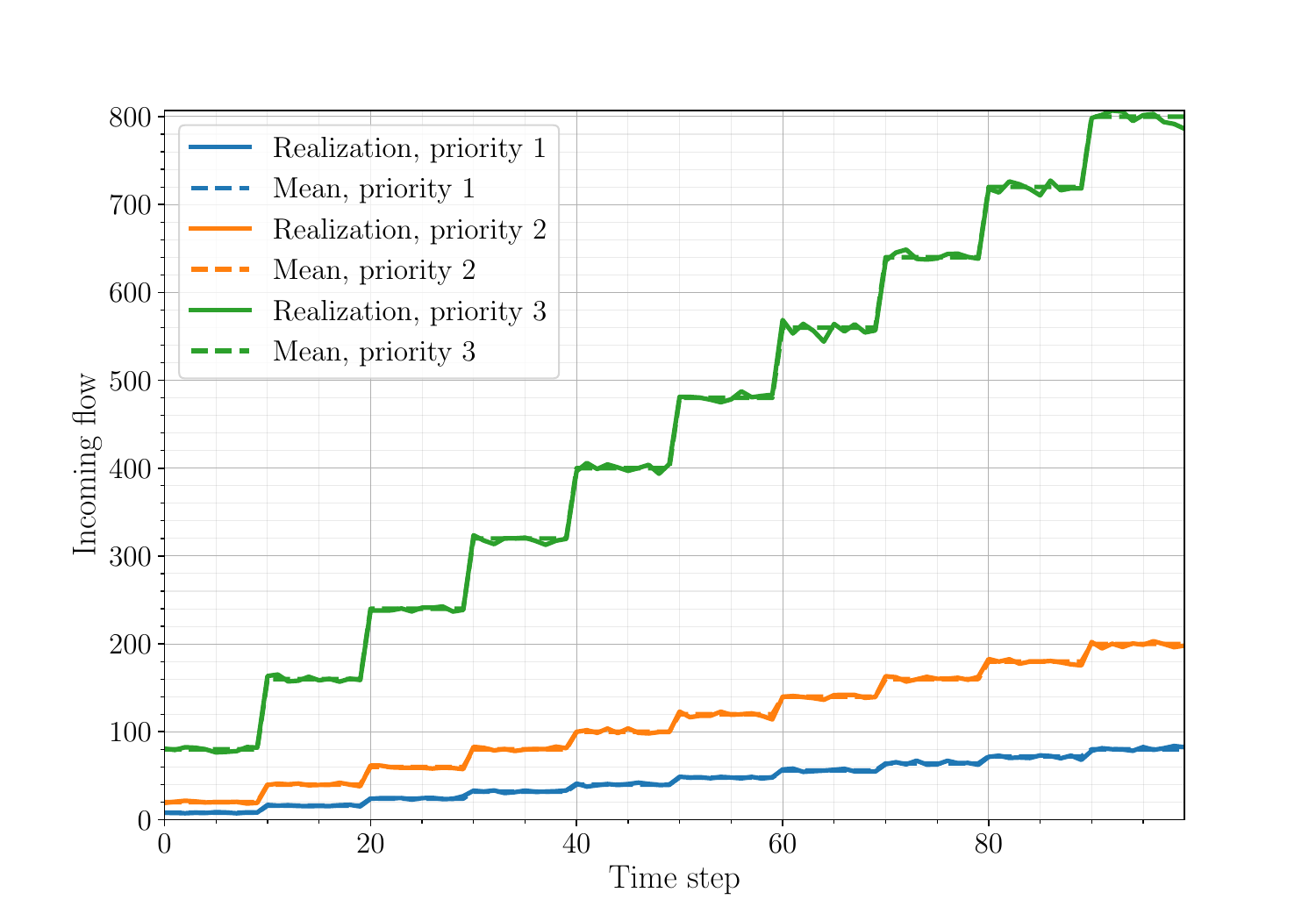}
        \caption{}
        \label{fig:f_in_vs_t_monte_carlo}
    \end{subfigure}
    \caption{Incoming flows across time (a) for a single run (b) averaged over 100 Monte-Carlo runs}
    \label{fig:f_in_vs_t}
\end{figure}

Initially, understanding the flow dynamics is essential as they form the basis of the system's operation. The incoming flow distribution is depicted in Figure \ref{fig:f_in_vs_t}. Figure \ref{fig:f_in_vs_t_single_run} displays the incoming flows for a single run, highlighting the system's inherent uncertainty. In contrast, Figure \ref{fig:f_in_vs_t_monte_carlo} showcases the incoming flows over time, averaged across 100 Monte-Carlo runs.

\begin{figure}
    \centering
    \includegraphics[width=\columnwidth]{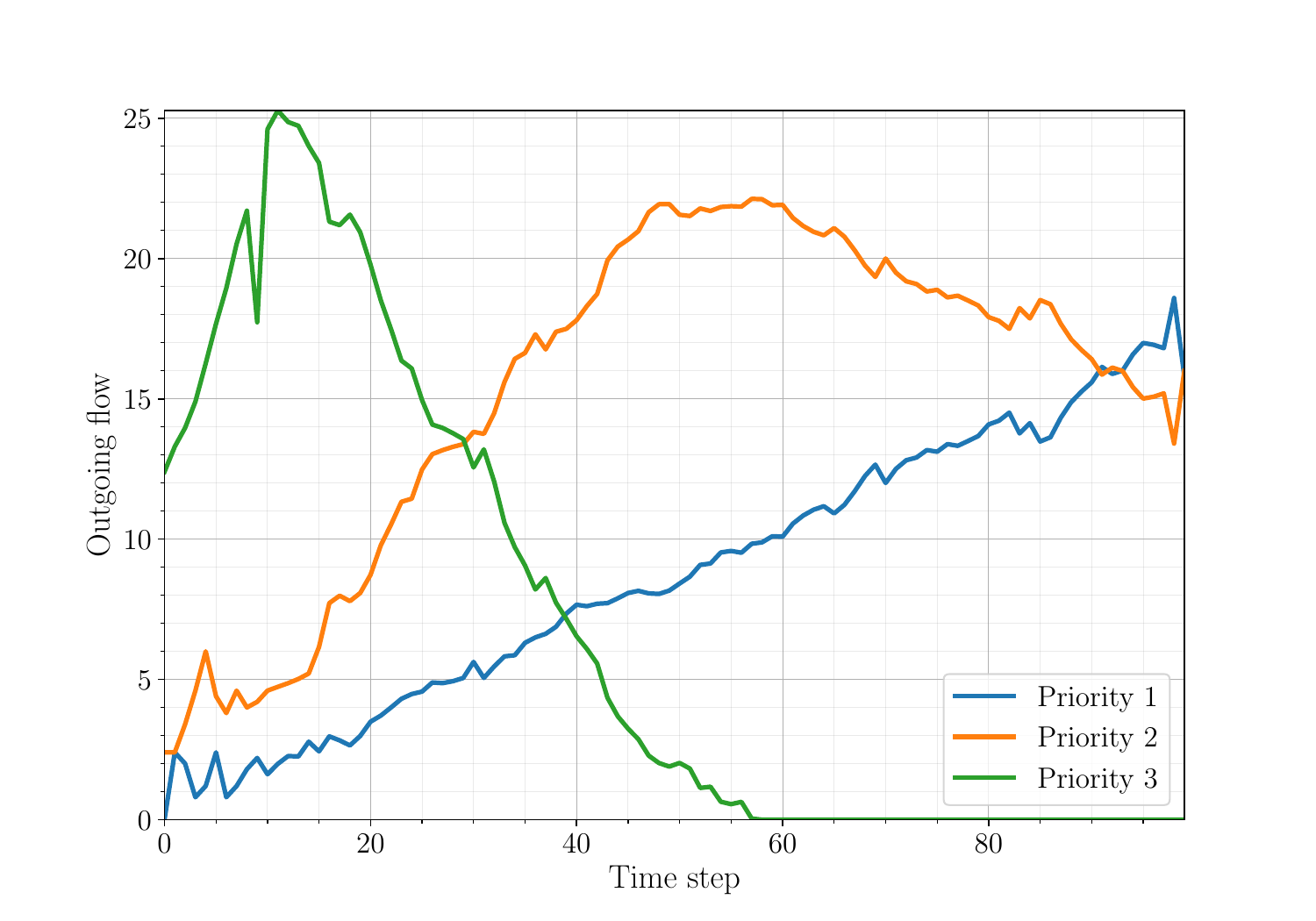}
    \caption{Outgoing flows across time averaged over 100 Monte-Carlo runs}
    \label{fig:f_out_vs_t}
\end{figure}

Next, we analyze the aggregated outflows across all $M$ modules for each priority level. The results, specific to our MPC approach, are illustrated in Figure \ref{fig:f_out_vs_t}.

As illustrated in Figure \ref{fig:f_out_vs_t}, when the system tends towards saturation, the MPC approach prioritizes transmitting packets of the highest priorities, discarding the lowest one.

\begin{figure}
    \centering
    \includegraphics[width=\columnwidth]{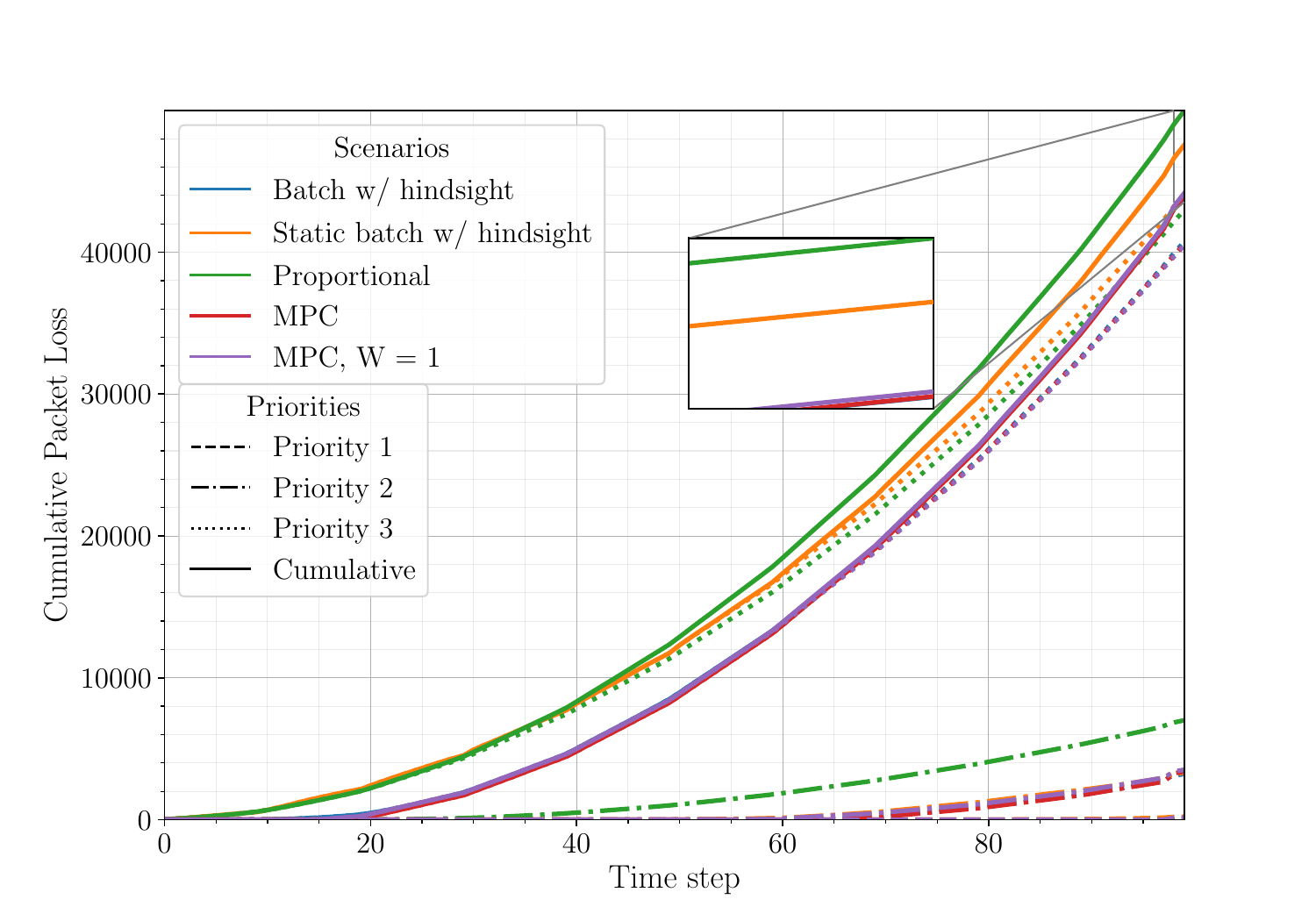}
    \caption{Lost packets across time averaged over 100 Monte-Carlo runs}
    \label{fig:L_vs_t}
\end{figure}

Figure \ref{fig:L_vs_t} presents the lost packets across time, delineated by priorities, for different scenarios. It shows that the \textit{batch optimization with hindsight} generally discards fewer packets of priority 1 than MPC but more of priorities 2 and 3, which led to a lower cost.

\begin{figure}
    \centering
    \includegraphics[width=\columnwidth]{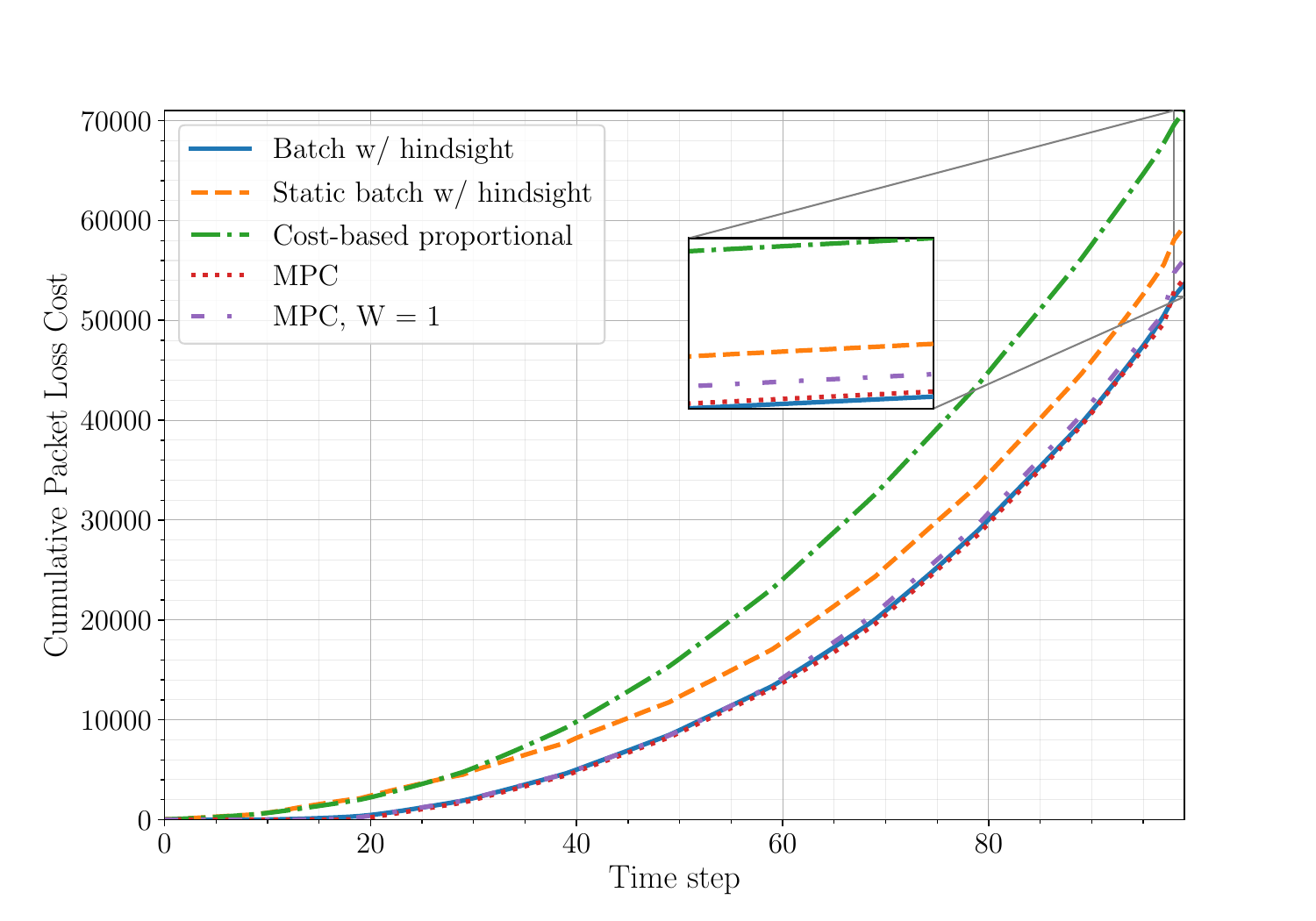}
    \caption{Packet loss cost across time averaged over 100 Monte-Carlo runs}
    \label{fig:L_cost_vs_t}
\end{figure}

Building on Figure \ref{fig:L_vs_t}, we now focus on the central aspect of this study: packet loss costs. Minimizing packet loss cost is the primary objective of these optimization efforts. Figure \ref{fig:L_cost_vs_t} illustrates the comparative performance of various methods in minimizing packet loss costs over time. The results align with initial expectations. The \textit{batch with hindsight} method unsurprisingly emerged as the best technique. As previously highlighted, this method leverages hindsight information about the actual realization of the stochastic process modeling the flows, providing a rationale for its superior performance. Following closely, our MPC approach demonstrated its capabilities, with cumulative costs only 1.05\% greater than the \textit{batch with hindsight} method. Considering that the MPC method operates with the expected information rather than comprehensive data, this marginal increase underscores its effectiveness. Moreover, losses from the MPC method were 3.54\% inferior to those of the \textit{windowless MPC} method, which registered 4.59\% more costs than the hindsight optimum. The \textit{static batch with hindsight} method engendered costs 10.78\% higher than its dynamic \textit{batch with hindsight} counterpart and 9.73\% higher than the MPC approach. In contrast, the \textit{proportional approach} registered the highest costs with a total of 32.35\% more than the hindsight optimum. The significant cost disparity between our approach and the \textit{proportional} method highlights the benefits of the MPC and of a comprehensive scheduling mechanism.
 
\begin{figure}[h!]
    \centering
    \includegraphics[width=\columnwidth]{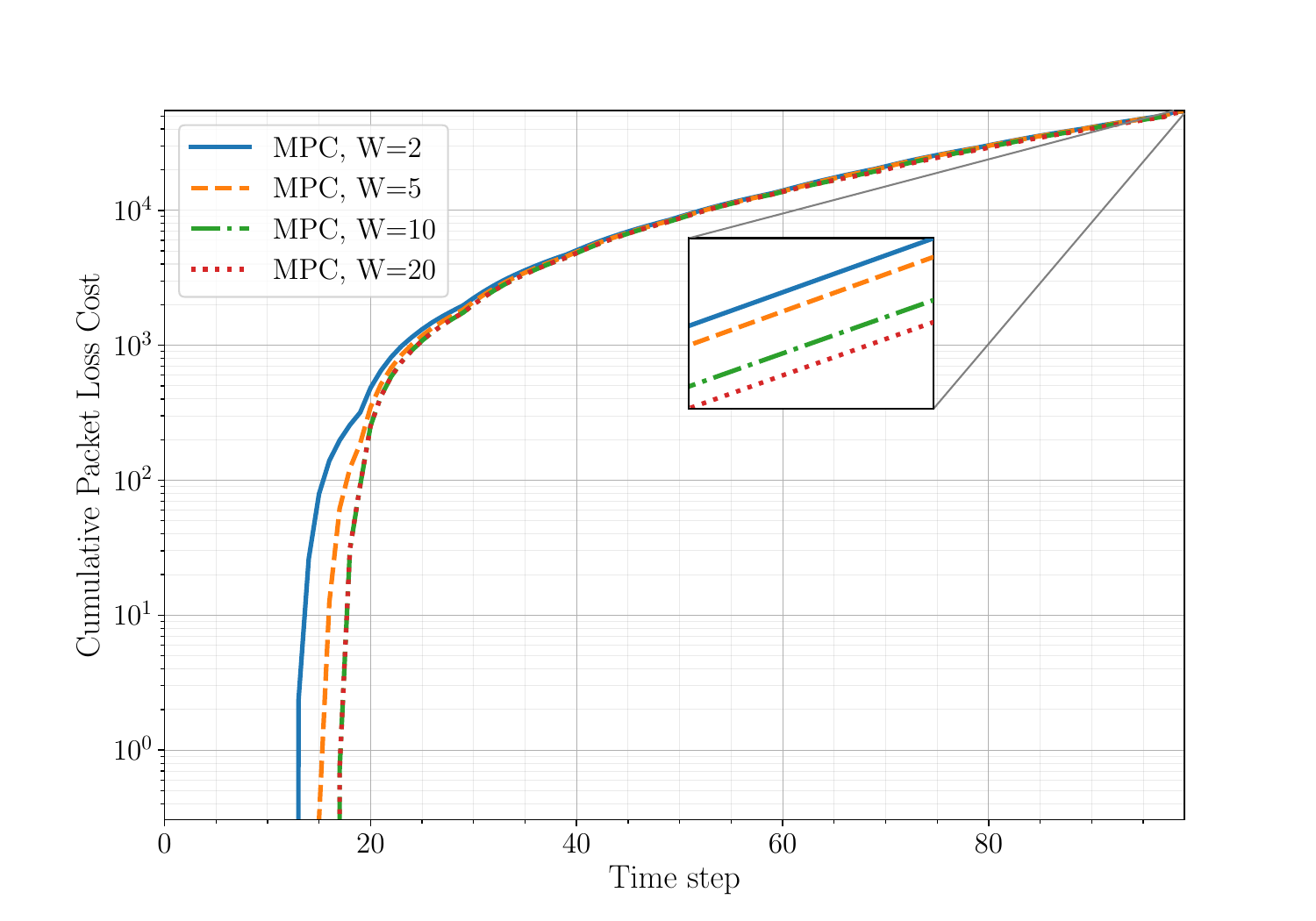}
    \caption{Packet loss cost across time averaged over 100 Monte-Carlo runs for different time windows}
    \label{fig:L_cost_vs_t_vs_W}
\end{figure}

Lastly, in our MPC method, we utilized a window size of 10. This choice proved to be advantageous, as it resulted in lower losses compared to smaller window sizes, as shown in Figure~\ref{fig:L_cost_vs_t_vs_W}. Specifically, the losses were 1.71\% and 1.19\% higher with window sizes of 2 and 5, respectively, when compared to the window size of 10. Additionally, it was found that a window size of 10 entails only 0.61\% more packet loss costs than a window size of 20, while requiring approximately half the computation time. This highlights the importance of balancing performance with computational efficiency, as the smaller window size of 10 achieves similar results to the window size of 20, but with significantly reduced computation time. These findings highlight the effectiveness of our chosen window size in optimizing the performance of our MPC approach.

\section{Conclusion}
This paper addresses an important challenge in satellite communications, focusing on the internal routing dynamics of HTS. As companies deploy an increasing number of satellites every year and as the world becomes more reliant on satellite communications for connectivity, ensuring efficient and adaptable packet allocation in HTSs is paramount.

The core contribution of this paper is the integration of an MPC technique to optimize the internal packet routing within HTSs. This method bridges the existing gap in the literature by offering a dynamic solution tailored to the uncertain environment of satellite communication. Although the \textit{batch optimization with hindsight} method is the optimal solution, its real-world applicability is limited because satellites never possess full time horizon information on the realization of the incoming flows. MPC strikes a balance between optimality and adaptability, making it more suitable for real-world deployment. The numerical results emphasize the potential of the MPC technique. Its performance, even though it only operates with mean information, is very close to the benchmark set by batch optimization. 

Future research will examine the scalability of MPC for larger systems and assess its computational efficiency on actual satellite hardware. Designing an online optimization-based strategy might offer improved adaptability to dynamic environments, prove beneficial for real-time applications, and represent a satellite-ready option.

\section*{Acknowledgments}
The authors thank MDA, the Consortium for Research and Innovation in Aerospace in Québec (CRIAQ), the
Institute for Data Valorization (IVADO), Mitacs, and the Natural Sciences and Engineering Research Council of
Canada (NSERC) for funding this project. 

\bibliographystyle{IEEEtran}
\bibliography{ref}

\begin{thebibliography}{10}
\providecommand{\url}[1]{#1}
\csname url@samestyle\endcsname
\providecommand{\newblock}{\relax}
\providecommand{\bibinfo}[2]{#2}
\providecommand{\BIBentrySTDinterwordspacing}{\spaceskip=0pt\relax}
\providecommand{\BIBentryALTinterwordstretchfactor}{4}
\providecommand{\BIBentryALTinterwordspacing}{\spaceskip=\fontdimen2\font plus
\BIBentryALTinterwordstretchfactor\fontdimen3\font minus \fontdimen4\font\relax}
\providecommand{\BIBforeignlanguage}[2]{{%
\expandafter\ifx\csname l@#1\endcsname\relax
\typeout{** WARNING: IEEEtran.bst: No hyphenation pattern has been}%
\typeout{** loaded for the language `#1'. Using the pattern for}%
\typeout{** the default language instead.}%
\else
\language=\csname l@#1\endcsname
\fi
#2}}
\providecommand{\BIBdecl}{\relax}
\BIBdecl

\bibitem{Richharia2010}
M.~Richharia and L.~D. Westbrook, \emph{Satellite Systems for Personal Applications: Concepts and Technology}.\hskip 1em plus 0.5em minus 0.4em\relax Wiley, Aug 2010.

\bibitem{IrekvistLinder2022}
\BIBentryALTinterwordspacing
J.~Irekvist and P.~Linder. {“Technology’s Important Role in Bridging Canada’s Digital Divide", 2022}. [Online]. Available: \url{https://www.ericsson.com/en/blog/6/2022/ericsson-bridging-canadas-digital-divide}
\BIBentrySTDinterwordspacing

\bibitem{pratt2019satellite}
T.~Pratt and J.~E. Allnutt, \emph{Satellite Communications}.\hskip 1em plus 0.5em minus 0.4em\relax John Wiley \& Sons, 2019.

\bibitem{Graydon2020}
M.~Graydon and L.~Parks, ``{'Connecting the Unconnected': A Critical Assessment of US Satellite Internet Services},'' \emph{Media, Culture \& Society}, vol.~42, no.~2, p. 260–276, 2020.

\bibitem{10068542}
D.~Tuzi, T.~Delamotte, and A.~Knopp, ``{Satellite Swarm-Based Antenna Arrays for 6G Direct-to-Cell Connectivity},'' \emph{IEEE Access}, vol.~11, pp. 36\,907--36\,928, 2023.

\bibitem{electronics8060683}
K.-H. Lee and K.~Y. Park, ``{Overall Design of Satellite Networks for Internet Services with QoS Support},'' \emph{Electronics}, vol.~8, no.~6, 2019.

\bibitem{LUGLIO2022109352}
M.~Luglio, S.~Romano, C.~Roseti, and F.~Zampognaro, ``{Satellite Multi-Beam Multicast Support for an Efficient Community-Based CDN},'' \emph{Computer Networks}, vol. 217, p. 109352, 2022.

\bibitem{Zhang_2023}
C.~Zhang, Y.~Zhou, P.~Qin, and Z.~Zhao, ``High-throughput satellite flexibility design and modeling,'' \emph{Journal of Physics: Conference Series}, vol. 2469, no.~1, p. 012006, Mar. 2023.

\bibitem{Kuiper2019}
K.~S. LLC, ``Application for {Authority} to {Launch} and {Operate} a {Non-Geostationary} {Satellite Orbit System} in {Ka-band Frequencies: Technical Appendix},'' Tech. Rep., 2019, {Technical Report}.

\bibitem{yahia2023evolution}
O.~B. Yahia, Z.~Garroussi, O.~Bélanger, B.~Sansò, J.-F. Frigon, S.~Martel, A.~Lesage-Landry, and G.~{Karabulut Kurt}, ``{Evolution of High Throughput Satellite Systems: Vision, Requirements, and Key Technologies},'' 2023, arXiv:2310.04389.

\bibitem{itu-r2019key-elements}
I.~T. Union, ``{Key Elements for Integration of Satellite Systems into Next Generation Access Technologies},'' International Telecommunication Union, Tech. Rep. ITU-R M.2460-0, Jul. 2019.

\bibitem{9861699}
M.~M. Azari, S.~Solanki, S.~Chatzinotas, O.~Kodheli, H.~Sallouha, A.~Colpaert, J.~F. Mendoza~Montoya, S.~Pollin, A.~Haqiqatnejad, A.~Mostaani, E.~Lagunas, and B.~Ottersten, ``{Evolution of Non-Terrestrial Networks From 5G to 6G: A Survey},'' \emph{IEEE Communications Surveys \& Tutorials}, vol.~24, no.~4, pp. 2633--2672, 2022.

\bibitem{7417202}
M.~Cello, M.~Marchese, and F.~Patrone, ``{HotSel: A Hot Spot Selection Algorithm for Internet Access in Rural Areas through Nanosatellite Networks},'' in \emph{IEEE Global Communications Conference (GLOBECOM)}, 2015, pp. 1--6.

\bibitem{6934560}
S.~Agnelli, P.~Feltz, P.-F. Griffiths, and D.~Roth, ``{Satellite's Role in the Penetration of Broadband Connectivity within the European Union},'' in \emph{7th Advanced Satellite Multimedia Systems Conference and the 13th Signal Processing for Space Communications Workshop (ASMS/SPSC)}, 2014, pp. 306--311.

\bibitem{electronics11213611}
K.~Suresh, A.~Alqahtani, T.~Rajasekaran, M.~S. Kumar, V.~Ranjith, R.~Kannadasan, N.~Alqahtani, and A.~A. Khan, ``{Enhanced Metaheuristic Algorithm-Based Load Balancing in a 5G Cloud Radio Access Network},'' \emph{Electronics}, vol.~11, no.~21, 2022.

\bibitem{9843265}
N.~Pachler, E.~F. Crawley, and B.~G. Cameron, ``{Beam-to-Satellite Scheduling for High Throughput Satellite Constellations Using Particle Swarm Optimization},'' in \emph{IEEE Aerospace Conference (AERO)}, 2022, pp. 1--9.

\bibitem{8002583}
Z.~Qu, G.~Zhang, H.~Cao, and J.~Xie, ``{LEO Satellite Constellation for Internet of Things},'' \emph{IEEE Access}, vol.~5, pp. 18\,391--18\,401, 2017.

\bibitem{9500740}
O.~Markovitz and M.~Segal, ``{Advanced Routing Algorithms for Low Orbit Satellite Constellations},'' in \emph{{IEEE International Conference on Communications}}, 2021, pp. 1--6.

\bibitem{ahuja1993network}
R.~K. Ahuja, T.~L. Magnanti, and J.~B. Orlin, \emph{{Network Flows: Theory, Algorithms, and Applications}}.\hskip 1em plus 0.5em minus 0.4em\relax Upper Saddle River, NJ, USA: Prentice Hall, 1993.

\bibitem{medhi2018network}
D.~Medhi and K.~Ramasamy, \emph{Network Routing: Algorithms, Protocols, and Architectures}, 2nd~ed.\hskip 1em plus 0.5em minus 0.4em\relax Morgan Kaufmann, 2018.

\bibitem{luenberger2008linear}
D.~G. Luenberger and Y.~Ye, \emph{Linear and Nonlinear Programming}.\hskip 1em plus 0.5em minus 0.4em\relax Springer Science \& Business Media, 2008.

\bibitem{rawlings2017mpc}
J.~B. Rawlings, D.~Q. Mayne, and M.~M. Diehl, \emph{Model Predictive Control: Theory, Computation, and Design}, 2nd~ed.\hskip 1em plus 0.5em minus 0.4em\relax Nob Hill Publishing, LLC, 2017.

\bibitem{mosek}
{MOSEK ApS}, \emph{The MOSEK Optimization Software}, 2023, version 10.1 [Software]. Available from \url{https://www.mosek.com/}.

\end{thebibliography}

\end{document}